\documentclass[11pt]{article}

\usepackage{color}

\usepackage{amsfonts,amssymb,amsthm}
\usepackage{amsmath,amssymb}
\usepackage{mathrsfs}
\usepackage{enumerate}
\usepackage{srcltx}
\usepackage[ansinew]{inputenc}
\usepackage{mathtools}
    \mathtoolsset{showonlyrefs}
    \mathtoolsset{centercolon}
\usepackage{graphicx}

\newcommand{\0}{\mathbf 0}

\newcommand{\hb}{\mathbf{h}}

\newcommand{\qb}{\mathbf{q}}

\renewcommand{\div}{\text{div}\,}

\def\Real{{\mathord{\rm{I \kern-.22em R}}}}

\begin{document}

\begin{center}

\end{center}

\newcommand\email[1]{\texttt{#1}}
\newcommand\at{:}

\begin{center}
 {\bf \Large
A revised exposition of the Green--Naghdi theory of heat propagation}
\end{center}
\medskip

\begin{center}
{\large  Swantje Bargmann$^{\star,\diamondsuit}$ \quad
        Antonino Favata$^\star$ \quad
        Paolo Podio-Guidugli$^\diamond$
}\end{center}

\begin{center}
 \noindent $^\star$ Institute of Continuum Mechanics and Materials Mechanic\\Hamburg University of Technology\footnote{Eißendorfer Str. 42
 21073 Hamburg,
 Germany.
  \\
 {\null} \quad \ Email:
 \begin{minipage}[t]{30em}
  \email{swantje.bargmann@tu-harburg.de} (S. Bargmann)\\
  \email{favata@ing.uniroma2.it} (A. Favata)\\
 \end{minipage}} 
 
  \vspace{0.5cm}
  
  \noindent $^\diamondsuit$ Institute of Materials Research, Helmholtz-Zentrum\footnote{Max-Planck-Straße 1,  21502 Geesthacht, Germany.} 
 
 \vspace{0.5cm}
  \noindent $^\diamond$Dipartimento di Ingegneria Civile e Ingegneria Informatica, \\Universit\`a di Roma TorVergata\footnote{Via Politecnico 1, 00133 Rome, Italy. \\
{\null} \quad \ Email:
\begin{minipage}[t]{30em}
\email{ppg@uniroma2.it} (P. Podio-Guidugli)
\end{minipage}}
\small
\end{center}
\medskip

\begin{abstract}
\noindent {\footnotesize We offer a revised exposition of the three types of heat-propagation theories proposed by Green and Naghdi, in the form they were given in 1992-3 \cite{Green:1992,Green:1993}. Those theories, which make use of the notion of thermal displacement and allow for heat waves, are at variance with the standard Fourier theory; they have attracted considerable interest, and have been applied in a number of disparate physical circumstances, where heat propagation is coupled with elasticity, viscous flows, etc. \cite{Straughan2011}. However, their derivation is not exempt from criticisms, that we here detail, in hopes of opening the way to reconsideration of old applications and  proposition of new ones.

\medskip

}
\end{abstract}

\section{Introduction}
\label{intro}

The nonconventional heat conduction theory by Green and Naghdi \cite{gree91,Green:1992,Green:1993} provides a general framework within which a wider range of thermal problems than within the standard theory have been modeled. The theory is subdivided into three different types of different generality: Type I, under special circumstances,  encompasses  the standard theory based on Fourier's constitutive prescription for the heat flux vector; Type II allows for propagation of thermal waves without internal energy dissipation; and Type III is meant to provide a framework to describe an even wider range of problems, many of which are given a reasoned exposition in a recent book \cite{Straughan2011}. In fact, Green and Naghdi developed their theories, whatever the type, within a thermodynamic framework that they regarded as robust enough to deal, in addition to heat propagation in a rigid conductor, with a number of coupled phenomenologies such as those covered by thermoelasticity \cite{Green:1993} or by the theory of thermoviscous fluids \cite{Green:1995a,Green:1996}. 

This paper is devoted to point out and amend the faulty arguments and moot points we found in  Green--Naghdi theories, none of them excluded. Since the reservations we have are of a general thermodynamic nature, we confine our discussion to heat propagation in rigid conductors. 
In Section 2 we present our understanding of the thermodynamic structure common to all three types of   Green--Naghdi theories. This section serves as an indispensable background for our type-by-type scrutiny, to come in Sections 3 to 5. Our main criticisms  are listed in the conclusive Section 6.




%
%
%
%
%
%
%
%
%

\section{The common structure of Green--Naghdi theories}

We begin by recapitulating the field equations used by Green and Naghdi, in the form those equations are given in \cite{Green:1993}, a few notational changes apart.

\subsection{Entropy balance}
At variance with the classical procedure of postulating balance of energy (the so-called \emph{First Law})  and imbalance of entropy (the \emph{Second Law}), Green and Naghdi base their approach on a statement of  entropy balance, a procedure they first introduced in  \cite{Green:1977} (the consistency of the latter approach with the former is demonstrated in  \cite{Green:1991b}). 
Their \emph{entropy balance} reads:
\begin{equation}\label{entropybal}
 \dot{\eta}= -\text{div}\boldsymbol{h}+s+\xi,
\end{equation}
with $\eta$, $s$ and $\xi$, respectively, the entropy, external entropy supply and internal entropy production, and with $\boldsymbol{h}$ the entropy influx vector (here and henceforth a superposed dot signifies differentiation with respect time $t$). They assume  that the \emph{energy inflow} $(\qb,r)$, where $\qb$ is the heat influx vector and $r$ is the external energy supply,  is proportional to the \emph{entropy inflow} $(\hb,s)$  via the absolute temperature $\theta$:
\begin{equation}\label{relations}
\boldsymbol{q}=\theta\boldsymbol{h}\quad\text{and}\quad\theta s=r,\quad\text{with}\quad \theta>0.
\end{equation}

\vskip 6pt
\noindent\emph{Remark.} The generality of this classic assumption has been questioned by M\"uller \cite{muel71,Mueller:1971,Mueller:1985} and others, e.g.\, Liu \cite{liu72,Liu:1996,Liu:2002,Liu:2008,Liu:2009}. To the best of our knowledge, Green and Naghdi simply ignored those  reservations, which however were considered briefly in \cite{Green:1971}. 
The issue was taken up in \cite{Bargmann:2007}  by Bargmann and Steinmann, whose developments and results were straightened and completed by us in our paper \cite{Bargmann:2012}. Both in \cite{Bargmann:2007} and in \cite{Bargmann:2012}, the   Green--Naghdi theory of Type III is considered within a M\"uller-Liu framework; among other things, it is shown that $\eqref{relations}_1$ holds if both $\boldsymbol{h}$ and $\boldsymbol{q}$ depend \emph{isotropically} on the state variables, whereas it does not hold if that dependence is only \emph{transversely isotropic}. That isotropy  guarantees  $\eqref{relations}_1$, whereas transverse isotropy does not, was first suggested by Liu \cite{Liu:2009}, for elastic bodies studied in a M\"uller-Liu framework. Recently, one of us has offered a simple argument to prove, in a standard thermodynamic setting, that influx proportionality holds (does not hold) if heat conduction is isotropic (transversely isotropic) \cite{PodioGuidugli:untitled}.
\vskip 6pt

\subsection{Energy balance, free energy, and the reduced entropy balance}
Multiplication by $\theta$ of Eq.~\eqref{entropybal} yields:
\begin{equation}\label{entropytheta}
 \theta\dot{\eta}= -\text{div}\left(\boldsymbol{q}\right)+\theta^{-1}\,\boldsymbol{q}\cdot\nabla\theta+\theta(s+\xi)=-\text{div}\boldsymbol{q}+  \theta^{-1}\,\boldsymbol{q}\cdot\nabla\theta+r+ \theta\xi.
\end{equation}
With this, the \emph{energy balance}
\begin{equation}\label{enbal}
 \dot\varepsilon=-\div\qb+  r,
\end{equation}
and the notion of \emph{Helmhholtz free energy} per unit mass
\begin{equation}\label{freeen}
\psi=\varepsilon-\theta\eta,
\end{equation}
Green and Naghdi arrive at the following \emph{reduced entropy equation}:
\begin{equation}\label{energyred}
 \dot{\psi}+\eta\dot{\theta}+\theta^{-1}\,\boldsymbol{q}\cdot\nabla\theta+ \theta\xi=0;
\end{equation}
%
here $\theta\xi$ is the \emph{internal dissipation}.

\subsection{The thermal displacement}
Relation \eqref{energyred} is the common point of departure of the constitutive developments in   Green--Naghdi Type I, Type II, and Type III, theories, to be recapitulated here below. In all those theories, \eqref{energyred} plays the same role as the \emph{reduced entropy inequality}
\begin{equation}\label{redentineq}
 \dot{\psi}+\eta\dot{\theta}+\theta^{-1}\,\boldsymbol{q}\cdot\nabla\theta\leq 0
\end{equation} 
in the standard Coleman--Noll procedure \cite{Coleman:1963}. The difference is, of course, that Green and Naghdi do not assume once and for all that the internal dissipation should never be negative. 
Another major difference with the standard treatment of heat conduction is that, in all their three types of theory, Green and Naghdi include among the state variables, either implicitly or explicitly, the \textit{thermal displacement} $\alpha$, that is, by definition, a time primitive of an \emph{empirical temperature} $T$:
\[
\dot\alpha:=T;
\]
furthermore, they take the absolute temperature $\theta$ to be an affine function of $T$. Hereafter, for simplicity, we assume that empirical and absolute temperature coincide:
\begin{equation}\label{temps}
\dot\alpha:=\theta.
\end{equation}
\vskip 6pt
\noindent\emph{Remark.} A short historical account of the fortunes of thermal displacement, a notion that was first introduced by Helmholtz \cite{Helmholtz:1884} in 1884, is found in the appendix of \cite{PodioGuidugli:2009}. While the formal role of this notion in postulating the basic balance laws of thermomechanics is clear \cite{PodioGuidugli:2009}, a physical interpretation is still wanted, especially if consistent with the statistical-mechanics interpretation for temperature.  
\vskip 6pt


\subsection{State spaces}
For a \emph{state}, we mean a list of independent state variables, whose instantaneous values at a space point determine the current local state of a material body during a thermodynamic process of interest. Green and Naghdi based their approach on the introduction of three state spaces, one for each of their three types of heat conduction theory. Unfortunately, they were not consistent with their choice of state
spaces: in their first paper \cite{gree91}, 
state spaces of Types II and III include the thermal displacement $\alpha$; not so in their follow-up papers \cite{Green:1992} and \cite{Green:1993}. We have not been able to find any comment of theirs about this important change in their views. 
 In the following, we adopt the lists of state variables  that are found in \cite{gree91}, namely,
\begin{equation}\label{states}
 \mathcal{S}_\text{I}=\{\theta,\nabla \theta\}=\{\dot{\alpha},\nabla \dot{\alpha}\},\quad
 \mathcal{S}_\text{II}=\{{\alpha}, \dot{\alpha},\nabla\alpha\},\quad
 \mathcal{S}_\text{III}=\{{\alpha},\dot{\alpha},\nabla\alpha,\nabla \dot{\alpha}\}
\end{equation}
(needless to say, each subscript points to the type of theory the state it individuates is chosen for). Note that, in view of \eqref{temps}, $\theta$ is included in all three notions of state. Note also that Type III is the most
inclusive theory, while Types I and II cannot be ordered with respect to
generality.

\subsection{Thermodynamic restrictions on response mappings}\label{sect_app}
Assume, as is customary, that all of the mappings for the fields in \eqref{energyred} that are in the need of constitutive specifications (namely, $\psi$, $\eta$,  $\boldsymbol{q}$, and $\xi$) depend in principle on one and the same chosen list of state variables $\mathcal S$: $\psi=\widehat\psi(\mathcal S)$, \ldots, $\xi=\widehat\xi(\mathcal S)$. Then,  \eqref{energyred} can be written as follows:
\begin{equation}\label{GNineq}
\partial_{\mathcal S}\widehat\psi({\mathcal S})\cdot \dot{\mathcal S}+\widehat\eta({\mathcal S})\dot{\theta}+\theta^{-1}\,\widehat{\boldsymbol{q}}({\mathcal S})\cdot\nabla\theta+ \theta\widehat\xi({\mathcal S})=0.
\end{equation}
Let us now choose $\mathcal S$ in \eqref{GNineq} to be $\mathcal S_\text{I}$. Then, granted the arbitrariness in the choice of a continuation $\dot{\mathcal S_\text{I}}dt$ for any process ending at ${\mathcal S_\text{I}}(t)$,  \eqref{GNineq} implies that:
\begin{equation}\label{listI}
\begin{aligned}
&\widehat\psi \;\;\textrm{can only depend on}\;\, \theta;\\
&\partial_{\theta}\widehat\psi(\theta)+\widehat\eta(\theta)=0;\\
&\theta^{-1}\,\widehat{\boldsymbol{q}}({\mathcal S}_\text{I})\cdot\nabla\theta+ \theta\widehat\xi({\mathcal S}_\text{I})=0.
\end{aligned}
\end{equation}
Furthermore, when ${\mathcal S}\equiv{\mathcal S}_\text{II}$, it so happens that ${\dot{\mathcal S}}_\text{II}\cap {\mathcal S}_\text{II}=\{\dot\alpha\}$; hence, \eqref{GNineq} yields:
\begin{equation}\label{listII}
\begin{aligned}
&\partial_{\theta}\widehat\psi({\mathcal S}_\text{II})+\widehat\eta({\mathcal S}_\text{II})=0;\\
&\partial_{\nabla\alpha}\widehat\psi({\mathcal S}_\text{II})+\theta^{-1}\,\widehat{\boldsymbol{q}}({\mathcal S}_\text{II})=\0;\\
&\partial_{\alpha}\widehat\psi({\mathcal S}_\text{II})+ \widehat\xi({\mathcal S}_\text{II})=0.
\end{aligned}
\end{equation}
Finally, when ${\mathcal S}\equiv{\mathcal S}_\text{III}$,  ${\dot{\mathcal S}}_\text{III}\cap {\mathcal S}_\text{III}=\{\dot\alpha,\nabla\dot\alpha\}$; this time, \eqref{GNineq} is satisfied identically (if and only) if
\begin{equation}\label{listIII}
\begin{aligned}
&\widehat\psi \;\;\textrm{can only depend on}\;\alpha,\,\theta,\,\textrm{and}\,\nabla\alpha;\\
&\partial_{\theta}\widehat\psi({\mathcal S}_\text{III})+\widehat\eta({\mathcal S}_\text{III})=0;\\
&\theta\,\partial_{\alpha}\widehat\psi({\mathcal S}_\text{III})+\big(\partial_{\nabla\alpha}\widehat\psi({\mathcal S}_\text{III})+\theta^{-1}\,\widehat{\boldsymbol{q}}({\mathcal S}_\text{III})\big)\cdot\nabla\theta+ \theta\widehat\xi({\mathcal S}_\text{III})=0.
\end{aligned}
\end{equation}
Note that the last relation in each of the three sets \eqref{listI}--\eqref{listIII} may be thought of as \emph{defining} the constitutive mapping that delivers the internal entropy production, in the following three forms:
\begin{equation}\label{csicost}
\begin{aligned}
\widehat\xi_\text{I}({\mathcal S}_\text{I})&=-\,\theta^{-2}\,\widehat{\boldsymbol{q}}_\text{I}({\mathcal S}_\text{I})\cdot\nabla\theta,\\
\widehat\xi_\text{II}({\mathcal S}_\text{II})&=-\,\partial_{\alpha}\widehat\psi_\text{II}({\mathcal S}_\text{II}),\\
\widehat\xi_\text{III}({\mathcal S}_\text{III})&=-\,\partial_{\alpha}\widehat\psi_\text{III}({\mathcal S}_\text{III})-\theta^{-1}\big(\partial_{\nabla\alpha}\widehat\psi_\text{III}({\mathcal S}_\text{III})+\theta^{-1}\,\widehat{\boldsymbol{q}}_\text{III}({\mathcal S}_\text{III})\big)\cdot\nabla\theta,
\end{aligned}
\end{equation}
that is, in terms of the mappings delivering the heat influx and/or the free energy.\footnote{Here, at variance with what we did  when writing \eqref{listI}--\eqref{listIII} to keep our notation lighter, we have carefully distinguished the heat-influx and free-energy constitutive mappings according to the type of theory they are meant for: e.g., we have written  $\widehat\psi_\text{I}$ in the case of Type I theory and $\widehat\psi_\text{II},\widehat\psi_\text{III}$  in the other two cases.} 
 
In the appendices to \cite{Green:1992} and \cite{Green:1993}, Green and Naghdi assert that, when \eqref{GNineq} is exploited to find restrictions on the constitutive mappings, the internal entropy production $\xi$ must be considered independent of the time rates of the state variables. This is indeed the case for their Type I theory, where $\dot{\mathcal{S}}_\text{I}\cap\mathcal{S}_\text{I}=\emptyset$, and they do arrive at the correct results \eqref{listI} in that case. But they regard that assertion about $\xi$ valid for Type II and Type III theories as well, whereas  both $\dot{\mathcal{S}}_\text{II}\cap\mathcal{S}_\text{II}$ and $\dot{\mathcal{S}}_\text{III}\cap\mathcal{S}_\text{III}$ are not empty; hence their line of reasoning is incorrect, as are their conclusions.
\vskip 6pt
\noindent\emph{Remark.} It is instructive to compare the results \eqref{listI}-\eqref{listIII} with the results that are obtained if one makes  about the internal entropy production, by far the most mysterious and least experimentally accessible of all the state functions, the minimally committing assumption that it  is nonnegative:
\begin{equation}\label{csi}
\xi\geq 0,
\end{equation}
an assumption that is at the conceptual core of standard continuum thermodynamics. A direct consequence of \eqref{csi} is that (inequality \eqref{energyred} takes the familiar form \eqref{redentineq}
and) inequality \eqref{GNineq} reads:
\begin{equation}\label{GNineqq}
\partial_{\mathcal S}\widehat\psi({\mathcal S})\cdot \dot{\mathcal S}+\widehat\eta({\mathcal S})\dot{\theta}+\theta^{-1}\,\widehat{\boldsymbol{q}}({\mathcal S})\cdot\nabla\theta\leq0.
\end{equation}
From \eqref{GNineqq}, in the case of Type I theory one derives $\eqref{listI}_{1,2}$ plus, in place of $\eqref{listI}_{3}$, the classic inequality for the heat influx:
\begin{equation}\label{fourier}
\widehat{\boldsymbol{q}}({\mathcal S}_\text{I})\cdot\nabla\theta\leq 0.
\end{equation}
Likewise, for Type II theory one re-obtains $\eqref{listII}_{1,2}$, while $\eqref{listII}_{3}$ is replaced by
\[
\partial_\alpha\widehat\psi({\mathcal S}_\text{II})\leq 0;
\]
and, for Type III, the only change is that $\eqref{listIII}_{3}$ is replaced by the residual entropy inequality
\[
\partial_{\alpha}\widehat\psi({\mathcal S}_\text{III}) + \theta^{-1}\big(\partial_{\nabla\alpha}\widehat\psi({\mathcal S}_\text{III})+\theta^{-1}\,\widehat{\boldsymbol{q}}({\mathcal S}_\text{III})\big)\cdot\nabla\theta\leq 0.
\]
\vskip 6pt
Green--Naghdi theories are given shape when the thermodynamic information in the lists from \eqref{listI} to \eqref{listIII} are inserted into the entropy balance \eqref{entropybal}. In all cases, the same equation is arrived at if the same information is inserted into the energy balance \eqref{enbal}, the procedure we adopt in the next three sections.

\section{Type I theory}
To begin with, we derive from definition \eqref{freeen}, with the use of
\[
\psi=\widehat\psi(\theta)\quad\textrm{and}\quad \widehat\eta(\theta)=-\partial_{\theta}\widehat\psi(\theta),
\]
the following general expression for the internal energy:
\[
\varepsilon= \widehat\psi(\theta)-\theta\,\partial_{\theta}\widehat\psi(\theta).
\]
Now, let us recall the classic prescriptions for free energy and heat influx:
\begin{equation}\label{eqcost}
\psi=-\lambda\theta(\log\theta -1)\quad\textrm{and}\quad \boldsymbol{q}=-\kappa\nabla\theta,
\end{equation}
where the \emph{heat capacity} $\lambda$ and the \emph{thermal conductivity} $\kappa$ are two positive constants (positivity of $\kappa$ is demanded for the Fourier law $\eqref{eqcost}_2$ to agree with the residual entropy inequality \eqref{fourier}). With $\eqref{eqcost}_1$, one finds:
\begin{equation}\label{etaeps}
\eta=\lambda\log\theta\quad\text{and}\quad\varepsilon=\lambda\,\theta;
\end{equation}
the end result is the well-known \emph{heat equation}:
\begin{equation}\label{he}
\lambda\dot\theta=\kappa\,\Delta\theta+ r,
\end{equation}

At variance with \eqref{eqcost}, Green and Naghdi assume that
\begin{equation}\label{constitutiveI}
\widehat\psi_\text{I}(\theta)=-\frac{1}{2}\lambda\frac{\theta^2}{\theta_0},
\end{equation}
and take
\begin{equation}\label{entropyfluxI}
\widehat{\boldsymbol{h}}_\text{I}(\theta,\nabla\theta)=-\frac{\kappa}{\theta_0}\nabla\theta,
\end{equation}
for the entropy influx, $\theta_0$ being a constant reference temperature;  \eqref{constitutiveI} and \eqref{entropyfluxI} imply that, respectively,
\begin{equation}\label{entI}
\widehat\eta_\text{I}(\theta)=\lambda\,\frac{\theta}{\theta_0}
\end{equation}
and
\begin{equation}\label{heatfl}
\widehat{\boldsymbol{q}}_\text{I}(\theta,\nabla\theta)=-\kappa\,\frac{\theta}{\theta_0}\,\nabla\theta.
\end{equation}
It follows from the above constitutive assumptions that
\begin{equation}\label{enI}
\widehat{\varepsilon}_{\text{I}}(\theta)=\frac{1}{2}\lambda\frac{\theta^2}{\theta_0};
\end{equation}
moreover, the first of \eqref{csicost} specifies the internal entropy production:
\[
{\widehat\xi}_\text{I}(\theta,\nabla\theta)=\frac{\kappa}{\theta_0}\,\theta^{-1}\,\nabla\theta\cdot\nabla\theta.
\]
Insertion of \eqref{enI} and \eqref{heatfl} into \eqref{enbal} yields, with the use of $\eqref{listI}_3$ and $\eqref{relations}_2$, a balance equation pretty different from the classic heat equation \eqref{he}, namely,
\[
\lambda\dot\theta=\kappa\,\Delta\theta+\theta_0(\xi_\text{I}+s).
\]
Green and Naghdi get to \eqref{he} by assuming that ``\ldots the temperature $\theta$ represents departure from an equilibrium temperature $\theta_0$ and \ldots time and space derivatives of $\theta$ \ldots are small of $O(\epsilon)$ \ldots'' (\cite{Green:1993}, p. 197), which is slightly more than needed to take
\begin{equation}\label{xxx}
\theta_0\,{\widehat\xi}_\text{I}(\theta,\nabla\theta)\simeq 0, \quad \theta_0 s\simeq r,
\end{equation}
and, said differently, amounts to restrict attention to almost isothermal and spatially uniform heat propagation processes, for which
\begin{equation}\label{const_mist}
\theta\simeq\theta_0\quad\textrm{and}\quad |\nabla\theta|\simeq 0.
\end{equation}

While one might be willing to study the nonlinear parabolic PDE Green and Naghdi arrive at before proceeding to their so-called `linearization', namely,
\[
\lambda\dot\theta=\kappa\,\theta^{-1}\div(\theta\,\nabla\theta)+\theta^{-1}\theta_0\,r,
\]
we see no point in deducing \eqref{he} through the involved path they chose.

\section{Type II theory}
We read in \cite{Green:1993} the following recipe for Type II free energy:
\begin{equation}\label{eq_psi2}
 \widehat\psi_{\text{II}}({\alpha}, \theta,\nabla\alpha)=-\frac{1}{2}\lambda\,\frac{\theta^2}{\theta_0}+\frac{1}{2}\frac{\kappa^\star}{\theta_0}\nabla\alpha\cdot\nabla\alpha\Big(= \widehat\psi_{\text{I}}(\theta)+\frac{1}{2}\frac{\kappa^\star}{\theta_0}\nabla\alpha\cdot\nabla\alpha\Big),
\end{equation}
where $\kappa^\star$ denotes a non-standard thermal conductivity coefficient. Note that, as a direct consequence of assuming that the free energy does not depend on the thermal displacement,  it follows from $\eqref{listII}_3$ that \emph{the internal entropy production is null}:
\begin{equation}\label{noprod}
\widehat\xi_{\text{II}}({\alpha}, \theta,\nabla\alpha)\equiv 0.\footnote{In our opinion, this  artifact should be considered less surprising than it seemed when \cite{gree91} and  \cite{Green:1993} appeared, because of the number of \emph{ad hoc} assumptions required to put it together.}
\end{equation}
The entropy associated with prescription \eqref{eq_psi2} for the free energy  turns out to be the same as in \eqref{entI}:
\begin{equation}\label{eq_entropy2}
 \widehat\eta_\text{II}({\alpha}, \theta,\nabla\alpha)=\lambda\,\frac{\theta}{\theta_0}=\widehat\eta_\text{I}(\theta);
\end{equation}
consequently,
\begin{equation}\label{eq_energy2}
\widehat{\varepsilon}_{\text{II}}({\alpha}, \theta,\nabla\alpha)=\frac{1}{2}\lambda\,\frac{\theta^2}{\theta_0}+\frac{1}{2}\frac{\kappa^\star}{\theta_0}\nabla\alpha\cdot\nabla\alpha.
\end{equation}
A peculiar property of Type II theory is that the energy and entropy influxes are both determined by the free energy, which plays the role of a potential function; we have from $\eqref{listII}_2$ that
\begin{equation}\label{eq_p2}
\widehat{\boldsymbol{q}}_{\text{II}}({\alpha}, \theta,\nabla\alpha)=-\frac{\kappa^\star}{\theta_0}\theta\,\nabla\alpha=\theta\,\widehat{\boldsymbol{h}}_{\text{II}}({\alpha}, \theta,\nabla\alpha).
\end{equation}
%
\vskip 6pt

\noindent\emph{Remark.} A well-known consequence of $\eqref{eqcost}_2$ is that the Fourier influx vector $\qb$ satisfies the classic \emph{heat-conduction inequality}
\[
\qb\cdot\nabla\theta\leq 0,
\]
which makes geometrically evident that heat influx and temperature gradient should always make an obtuse angle at any given body point in any given process of thermal propagation, in accord with common experience. One may ask whether the energy influx vector $\qb_{\text{II}}$ in \eqref{eq_p2} would satisfy the same inequality.
%
As a straightforward computation shows,
\[
\widehat{\boldsymbol{q}}_{\text{II}}({\alpha}, \theta,\nabla\alpha)\cdot\nabla\theta=-\frac{\kappa^\star}{2\theta_0}\theta\,\frac{d{}}{dt}|\nabla\alpha|^2;
\]
hence, provided that one assumes that $\kappa^\star\geq 0$, in a Type II theory the energy influx and temperature gradient form an obtuse angle only when $|\nabla\alpha|^2$ grows bigger. 
\vskip 6pt

Under the current assumptions, on making use of \eqref{noprod}, \eqref{eq_entropy2}, \eqref{eq_energy2}, and $\eqref{eq_p2}_2$, both the entropy balance \eqref{entropybal} and the energy balance \eqref{enbal} become:
\[
\lambda\dot\theta= {\kappa}^\star\Delta\alpha+{\theta}_0\, s,
\]
On recalling that $\dot{\theta}=\ddot{\alpha}$, the previous equation looks as follows:
\begin{equation}\label{nn}
\lambda\ddot\alpha= {\kappa}^\star\Delta\alpha+{\theta}_0\,{s},
\end{equation}
a hyperbolic PDE for $\alpha$. Green and Naghdi do not follow this path in \cite{Green:1993}: they rather differentiate \eqref{nn} with respect to time, presumably in order to get a PDE for temperature, a more familiar unknown field that thermal displacement, and a field for which assigning initial and boundary conditions is no problem. Following their suggestion, we get a hyperbolic PDE for $\theta$ that opens the way to study \emph{thermal wave propagation}, namely,
\begin{equation}\label{n3n}
\lambda\ddot\theta= {\kappa}^\star\Delta\theta+{\theta}_0\,\dot{ s}.
\end{equation}
Unfortunately, the equation Green and Naghdi arrive at in \cite{Green:1993} is:
\begin{equation}\label{eq_heat2}
\lambda\ddot{\theta}={\kappa}^\star\Delta\theta+ \dot{ r},\footnote{Most probably because of a typographical error, the dot signaling time differentiation of the last term is missing in \cite{Green:1993}.}
\end{equation}
an equation which coincides with \eqref{n3n} only if one takes
\[
r\simeq \theta_0\,s,
\]
that is, only if attention is restricted to almost isothermal processes, as was done in the case of Type I theory  by stipulating that
\eqref{const_mist}$_1$ hold.

\section{Type III theory}
Among the three types of heat propagation theories Green and Naghdi proposed, this is in our opinion the one whose deduction is, so to speak, the most adventurous. Here is how we reconstruct it, to the extent that it is possible,  from \cite{Green:1992}, the paper where Green-Naghdi present their Type III thermoelastic theory. 

The Type III free-energy mapping is the same as for a theory of Type II:
\begin{equation}\label{7777}
\widehat{\psi}_\text{III}(\alpha,\theta,\nabla\alpha)=-\lambda\,\frac{\theta^2}{2\theta_0}+\frac{\kappa^\star}{2\theta_0}\nabla\alpha\cdot\nabla\alpha.
\end{equation}
As to energy and entropy influxes, it is proposed to take
\begin{equation}\label{heatflux3}
\widehat{\boldsymbol{q}}_\text{III}(\alpha,\theta,\nabla\alpha,\nabla\theta)=-\frac{\kappa^\star+\kappa^{\star\star}}{\theta_0}\,\theta\,\nabla\alpha-\frac{\kappa}{\theta_0}\,\theta\,\nabla\theta
=\theta\,\widehat{\boldsymbol{h}}_\text{III}(\alpha,\theta,\nabla\alpha,\nabla\theta).
\end{equation}
Note that
\[
\widehat{\boldsymbol{q}}_\text{III}(\alpha,\theta,\nabla\alpha,\nabla\theta)=\widehat{\boldsymbol{q}}_\text{I}(\theta,\nabla\theta)+
\widehat{\boldsymbol{q}}_{\text{II}}({\alpha}, \theta,\nabla\alpha)-\frac{\kappa^{\star\star}}{\theta_0}\,\theta\,\nabla\alpha,
\]
so that, in addition to $\kappa$ and $\kappa^\star$, another thermal conductivity modulus -- but not another form of heat conduction! -- enters the theory; presumably, $\kappa^{\star\star}\geq 0$.
In view of $\eqref{listIII}_2$, and as a consequence of \eqref{7777} and \eqref{heatflux3}, we have that
\begin{equation}\label{eq_rela33}
\widehat{\eta}_\text{III}(\alpha,\theta,\nabla\alpha)= \lambda\, \frac{\theta}{\theta_0},
\end{equation}
so that, on recalling \eqref{entI} and \eqref{eq_entropy2}, entropy, whatever the type of theory, depends only on temperature, and in a manner that can be regarded as a first approximation for $\theta\simeq\theta_0$ of the classic prescription $\eqref{etaeps}_1$. Moreover, as a consequence of \eqref{7777} and \eqref{heatflux3}, we have that
\begin{equation}\label{eq_rela3}
\widehat{\xi}_\text{III}(\alpha,\theta,\nabla\alpha,\nabla\theta)=\theta^{-1}\Big(\frac{\kappa^{\star\star}}{\theta_0}\nabla\alpha+\frac{\kappa}{\theta_0}\nabla\theta\Big)\cdot\nabla\theta,
\end{equation}
or rather, equivalently,
\begin{equation}\label{eq_rela34}
\widehat{\xi}_\text{III}(\alpha,\theta,\nabla\alpha,\nabla\theta)=\theta^{-1}\Big(\frac{\kappa^{\star\star}}{\theta_0}\frac{d{}}{dt}(\nabla\alpha)^2+\frac{\kappa}{\theta_0}|\nabla\theta|^2\Big).
\end{equation}
Thus, once again, the expression for a quantity that should never be negative -- this time, the internal entropy production -- turns out to include a term that, just as the first one between brackets, could instead induce  negativity during some process.

But this is not all. Green and Naghdi \cite{Green:1992} claim that substitution of  Eqs.~\eqref{eq_rela33}, $\eqref{heatflux3}_2$ and \eqref{eq_rela3} 
into the `linearized' form of what they refer to as `Eq.~(2.17)' would yield:
\begin{equation}\label{heat_eq3}
\lambda\ddot{\theta}=\kappa\,\Delta\dot{\theta}+\kappa^\star\Delta\theta+\dot{r}.
\end{equation}
Now, `Eq.~(2.17)' is not found in \cite{Green:1992}. If, imitating the procedure followed  in \cite{gree91,Green:1993}, we perform the same substitution in the entropy balance \eqref{entropybal}, we find:
\begin{equation}
\lambda\dot{\theta}=\kappa\Delta{\theta}+\kappa^\star\Delta\alpha+\theta_0 s+\kappa^{\star\star}\Delta\alpha+\theta_0\widehat{\xi}_\text{III}(\alpha,\theta,\nabla\alpha,\nabla\theta),
\end{equation} 
or rather, on accepting again the pseudolinearization $\eqref{const_mist}$,
\[
\lambda\dot{\theta}\simeq \kappa\Delta{\theta}+\kappa^\star\Delta\alpha+r+\kappa^{\star\star}(\Delta\alpha+\nabla\alpha\cdot\nabla\theta).
\]
We fail to see how one could get rid of the last two terms on the right side in a manner both convincing and different from taking $\kappa^{\star\star}=0$. But then, one could infer that, for Type III theory, the internal entropy production is approximately null:
\begin{equation}\label{csiappr}
\widehat{\xi}_\text{III}(\alpha,\theta,\nabla\alpha,\nabla\theta)\simeq 0.
\end{equation}


%


\section{Conclusions}
In a number of papers appeared between 1977 and 1996 \cite{Green:1977,Green:1991b,gree91,Green:1992,Green:1993,Green:1995a,Green:1996}, Green and Naghdi have proposed three types of heat propagation theory. According to one of these theories, heat propagates without internal entropy production; in two of them, heat waves are possible. For these and other reasons, all Green-Naghdi theories, and especially so Type III theory, have soon attracted an interest that does not seem to fade away even at the time of this writing, and have been applied in a number of disparate physical circumstances, where propagation of heat is coupled with elastic deformations of solids, flow of viscoelastic fluids, etc. (an updated account of the applications we allude at is found in \cite{Straughan2011}).  Unfortunately, in our opinion, Green--Naghdi theories have the severe limitations pointed out in our present exposition; the three main ones are recapitulated hereafter. 
\begin{itemize}
\item The internal entropy dissipation $\xi$ is presumed to be the object of a constitutive prescription, which is in terms of the same state variables  as Helmholtz free energy $\psi$ and energy and entropy influxes $\boldsymbol{q},\boldsymbol{h}$, and turns out to depend on their constitutive prescriptions as specified by \eqref{csicost}; in particular, $\xi$ must be null in all processes considered within the framework of Type II theory. Now, a fundamental tenet of continuum thermodynamics, that was not made part of Green-Naghdi theories neither by their creators nor by their followers, is that $\xi$ \emph{must be non negative}. To conform to this tenet for Type III theory, one should be able to \emph{prove} that $\xi$ is non negative along whatever process, something that turns out to be possible under assumptions that would kind of emasculate that theory, at least partly.\footnote{In fact, it would be necessary to take $\kappa\geq 0$ (a reasonable assumption) and $\kappa^{\star\star}=0$, an assumption that he have seen to be expedient also to obtain the `heat equation' \eqref{heat_eq3}, and that, together with $\eqref{const_mist}_2$, would yield \eqref{csiappr}.}
\item The energy and entropy influx vectors are taken proportional via the temperature ($\boldsymbol{q}=\theta\boldsymbol{h}$), an assumption that has been proven to hold only when heat propagation is, in a precise sense, isotropic  \cite{Bargmann:2012}.
\item The notion of thermal displacement plays a rather ambiguous role as a state variable, in that: (i) it does not enter the constitutive prescriptions \emph{per se}, but only through its time and space derivatives $\dot\alpha,\nabla\alpha$ and $\nabla\dot\alpha$; (ii) as a exemplified by \eqref{nn} and \eqref{eq_heat2},  the direct outcome of a theory, a PDE for $\alpha$, is quickly differentiated with respect to time to arrive at a PDE for the temperature $\theta$, a field variable for which is indeed natural to lay down initial and boundary conditions (we wonder how one could be confident that an assignment of $\alpha$ at an initial time or on part of the boundary could be reproduced in the laboratory.) 
\end{itemize}


\noindent In spite of these criticisms, we believe that Green--Naghdi
heat conduction theories, when amended as indicated, are worth studying and applying, especially if a convincing statistical-mechanics interpretation is found for the thermal displacement. Thermal wave propagation is an example of  worth-pursuing research theme: it would be interesting to compare the predictions, say, of equation \eqref{heat_eq3} with those of other mathematically similar equations that have been proposed (for an account of the literature up to 1990 see \cite{Joseph:1989,Joseph:1990}; see also \cite{Naghy:1994}, where other references are found, and a very recent contribution of one of us \cite{PodioGuidugli:2012}).



\scriptsize{
\section*{Acknowledgments}
This research was done while PPG was visiting Hamburg University of Technology and Helmholtz-Zentrum Geesthacht. The financial support of the German Science Foundation is gratefully acknowledged.
}


\end{document}